\documentclass{mem}
\usepackage{natbib}\usepackage{txfonts}\usepackage{balance}
\usepackage{graphicx}
\usepackage[a4paper]{hyperref}
\idline{75}{282}

\begin{document}
\def\teff{$T\rm_{eff }$}
\def\kms{$\mathrm {km s}^{-1}$}

\title{
Shocking News outside of Cluster Cores
}

   \subtitle{}

\author{
L. \,Rudnick\inst{1},
S. \,Brown\inst{2}
\and D. \,Farnsworth\inst{1}
          }

 % \offprints{L. Rudnick}

\institute{
University of Minnesota, Department of Astronomy ,
116 Church St. SE,
Minneapolis, MN, 55455, USA.  
\email{larry@umn.edu, farnsworth@astro.umn.edu}
\and
CSIRO Astronomy \& Space Science,
Australian Telescope National Facility,
PO Box 76,
Epping NSW 1710,
Australia. \email{Shea.Brown@csiro.au}
}

\authorrunning{Rudnick }

\titlerunning{Shocking News}

\abstract{
Below $\sim$1$\mu$G, the minimum pressures of synchrotron plasmas start to approach those in the thermal gas in cluster outskirts and the more diffuse WHIM in large-scale structure filaments.  We summarize some of our techniques to find the corresponding low surface brightness radio sources and what he have uncovered.   We identify cluster-like sources in poor environments, likely requiring very efficient relativistic particle acceleration, as well as very diffuse radio galaxies with no indications of current activity.  More detailed observations of Coma and a new radio-identified cluster highlight some emerging issues in cluster-related shocks:   Coma has a shock  on the border of its halo, suggesting a connection between the two;  the Coma relic appears to be in ``infall" shock, associated with a 500 km/s infalling column of galaxies;  the new cluster has filamentary pieces in addition to its more classical halo and peripheral relics, suggesting that we are beginning to see relic emission face on.  Finally, a different kind of shock comes from our recent finding of ambiguities in the determination of rotation measures  which will influence cluster magnetic field studies. }

\maketitle{}

\section{Searching for low surface brightness emission}

Beyond cluster cores, in the infall regions and along the connecting filaments, we expect to find rich dynamical structures, as seen in numerical simulations that include the baryonic components.  These regions are likely to be permeated by stronger (higher Mach number) shocks than in cluster cores, although the diffuse baryons are difficult to detect.  Here, radio emission can provide unique diagnostics of the cluster formation process and the energization of magnetic fields and relativistic particles.   Since overpressured radio plasmas would expand very rapidly, we therefore search for low brightness emission as the indicator of low minimum pressure conditions.

\begin{figure*}[t!]
\resizebox{\hsize}{!}{\includegraphics[clip=true]{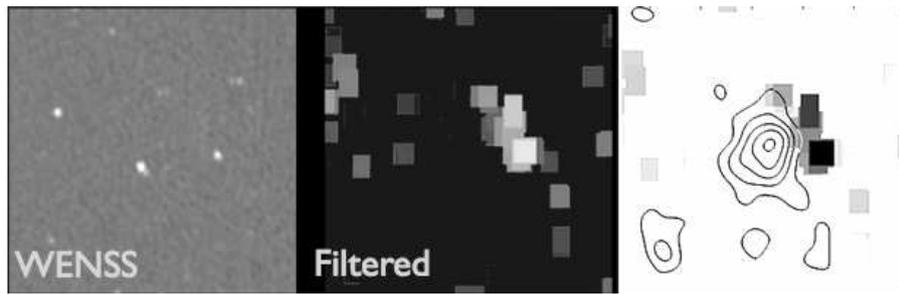}}
\caption{\footnotesize
Example of multi-resolution filtering to identify low surface brightness emission in existing surveys.  Left: original WENSS survey image.  Middle: image of low surface brightness emission with compact sources removed.  Right: Overlay of low brightness radio emission from RXJ1053.7+5450 on X-ray contours from archival ROSAT broadband image, convolved to 240\arcsec. (See \cite{rudlem}).
}\label{filter}
\end{figure*}

Detection of low surface brightness emission is challenging;  to reach required sensitivities we have to average over large areas, which rapidly become  "confused" by more compact, distant radio galaxies.  We have been developing techniques to remove compact source confusion in existing large surveys, 

{\em Multi-resolution filtering} was introduced by \cite{rudnick02}, based on an algorithm described in \cite{starck}.  It consists of a very fast method to decompose an image into any number of component images each respresenting a different angular resolution.  We illustrate it here in Figure \ref{filter} in its two component mode, where it is used to remove the point source contributions from the WENSS survey \citep{WENSS}.  This allows detection of much lower surface brightness emission than would otherwise be practical.  Figure \ref{filter} shows a new peripheral relic discovered around the X-ray cluster RXJ1053.7+5450 \cite{rudlem}.   A variety of cluster and non-cluster diffuse sources have been found in this way;  one interesting example is discussed below.

{\em NVSS polarization reprocessing.}
Nominally, the NVSS survey \citep{NVSS} is not sensitive to structures on scales greater than $\sim$15\arcmin.  However, even structures on the scale of 10s of degrees may be detectable in polarization, if their Q and U Stokes parameters vary on smaller scales.  Such variations are, in fact, common due to differential Faraday rotation and orientation of the magnetic field in the plane of the sky.  
\begin{figure}[]
\resizebox{\hsize}{!}{\includegraphics[clip=true]{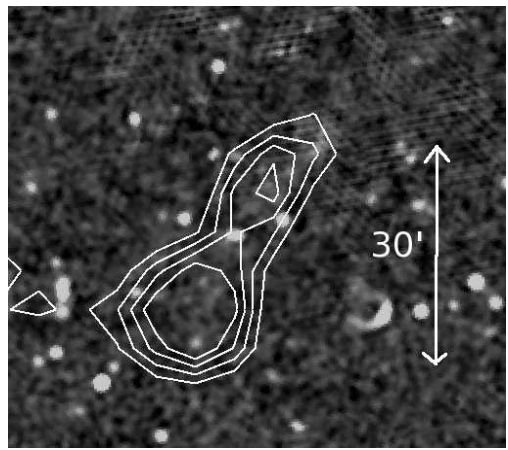}}
\caption{
\footnotesize
NVSS  (1.4~GHz) total intensity in greyscale,  SW of the Coma Cluster in the region of the relic.  Only hints of the relic are visible in the NVSS total intensity images.  Contours show the reprocessed NVSS polarization data, with the bright southern half of the relic prominently seen.  
}
\label{polar}
\end{figure}
The problem in detecting such emission, however, is that one cannot average Q and U over large angles to recover very low surface brightness emission.   In \cite{rudnickpol} we introduced a technique for recovering very large scale low surface brightness structures through their Q, U variations.  We averaged $<P>=<\sqrt(Q^2+U^2)>, $ a procedure normally discouraged in polarization work.  In this way, we were able to recover the polarized emission from our Galaxy over many 10s of degrees, as well as a wide variety of galactic and extragalactic sources.  One example is shown in Figure \ref{polar}, the relic to the southwest of the Coma cluster, which is marginally visible in the NVSS total intensity image, but seen at high signal:noise in our polarization reprocessing.
\begin{figure}[]
\resizebox{\hsize}{!}{\includegraphics[clip=true]{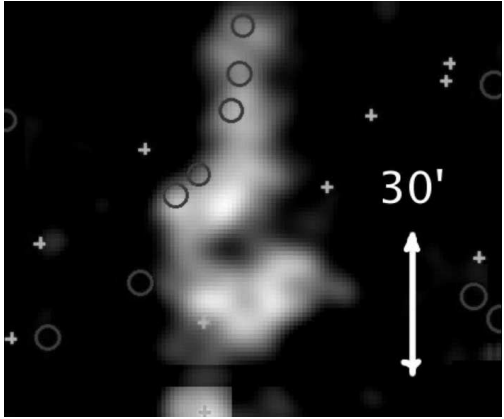}}
\caption{
\footnotesize
Extended feature seen in reprocessed NVSS polarization data  due to a Faraday screen, not a separate synchrotron source \citep{rudnickpol}.
}
\label{polar2}
\end{figure}
We have been following up on a number of the newly detected sources on the WSRT and the GBT, to disappointing results.  One example is shown in Figure \ref{polar2}.  This dramatic looking feature is not detected in total intensity at WSRT (330~MHz), but is instead visible as a patchy distribution of rotation measures to the diffuse galactic background.   Such a patchy RM distribution would produce small-scale variations in Q and U in the NVSS images, and show up as a ``pseudo-source" in our reprocessed polarization images.

\subsection{Cluster-like emission in poor environments}

One of the discoveries from our filtering procedure is a pair of unrelated diffuse sources at 0809+30 \citep{delain}.  In fact, these are strong enough to be visible in the original WENSS images.  Detailed mapping of these with both the WSRT and the VLA revealed them to be steep-spectrum objects with little or no associated X-ray emission \citep{brown0809}.  The southern of the two sources, S$_{diff}$ is associated with a chain of optical galaxies at a redshift of 0.04.   It is not clear whether it originated from a single AGN or is the result of a group-wide process.  If the latter, it is hard to see how there could be sufficient turbulence or shocks to accelerate fields and particles in this environment.

\begin{figure}[]
\resizebox{\hsize}{!}{\includegraphics[clip=true]{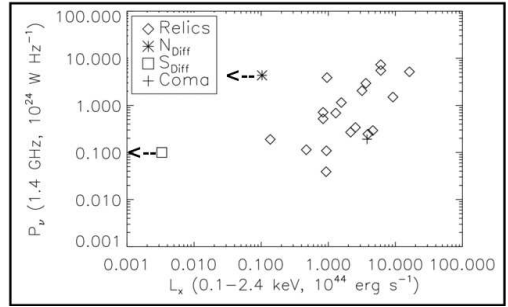}}
\caption{
\footnotesize
Plot of radio luminosity vs. X-ray luminosity for cluster relics, from \cite{brown0809}.  The two components of 0809+39 are shown as N$_{diff}$ and S$_{diff}$, with upper limits for their X-ray luminosities.  They are at least two orders of magnitude more radio luminous than expected.
}
\label{upper0809}
\end{figure}

The identification -- but not the physics -- of the northern of the two 0809+39 sources (N$_{diff}$) appear clearer.  It is apparently associated with a poor cluster of galaxies at a redshift of 0.2, with a wide-angle-tail \citep{owenrud} in the center.  In location, size, morphology, luminosity, spectrum and polarization it appears to be a quite ordinary relic, except for the lack of detectable X-ray emission from the cluster.

The relationship between radio and X-ray luminosities for these two sources are shown in Figure \ref{upper0809}.  They are approximately two orders of magnitude higher in radio luminosity than would be expected by comparison, e.g., with other cluster relic sources.  A filamentary structure in (projected on?) Abell 1213 \cite{a1213} also appears too strong for its X-ray luminosity.  In \cite{brown0809} we frame this dilemma in terms of the acceleration efficiency for relativistic particles.

At present, such objects are fairly rare.   We have followed up on several more of the filtering discoveries.   Two of these are very low surface brightness likely dead radio galaxies in modest galaxy groups (publication in preparation).  A third is an X-ray cluster source, discussed in the following section.   As the next generation of radio telescopes and surveys gets underway, We will likely see increasing numbers of ``disembodied" diffuse plasmas;  with luck they will help us illuminate the processes by which cluster relativistic plasmas originate and are re-energized.

\section{The cluster shock menagerie} 

As the number of observed halo and relic-type sources increases, and the distinctions between them blur, we have the opportunity to redefine our notion of shocks and turbulence in these very dynamic regions.  We need to look again to the increasingly 
sophisticated numerical simulations that include thermal and relativistic plasmas.  \cite{battaglia}, e.g., show us  a network of shocks inside the cluster, generated by previous accretion, mergers and turbulence.  These shocks are typically weak \citep{miniati} because of the high temperatures of the thermal plasma.  The various types of shocks we expect to eventually see associated with cluster and large scale structure formation include:
\begin{itemize}
\item{Internal cluster shocks --  may play a role in keeping halos energized;}
\item{Outgoing merger shocks, e.g., generating peripheral relics, \citep{relic};}
\item{Accretion shocks -- this term applies in the literature to the approximately spherical shocks due to continuing infall into the cluster.  They typically appear beyond the virial radius \citep{virial} and have {\em not} yet been seen in the radio;}
\item{Infall shocks -- a term we introduce in \cite{browncoma} to describe the shock at the termination of a column of galaxies infalling into Coma;}
\item{Filament shocks --  accompanying infall into and along filaments during large-scale structure formation \citep{ryu}.}
\end{itemize}
As we seek to disentangle the roles of these shocks, e.g., for relativistic particle acceleration,  it is important that we identify them properly and assign them the correct dynamical features.  The simple picture of outgoing merger shocks is  becoming inadequate for the rich variety of structures now seen.

\subsection{Revisiting the Coma cluster}

\begin{figure}[]
\resizebox{\hsize}{!}{\includegraphics[clip=true]{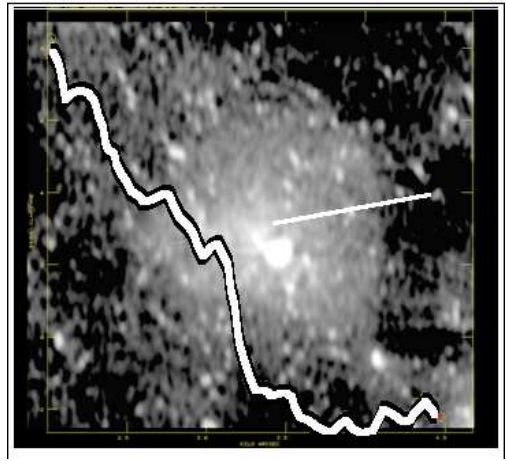}}
\caption{
\footnotesize
WSRT 330 MHz observations of Coma cluster; 1D cut showing shock at edge of halo \citep{browncoma}.
}
\label{comaedge}
\end{figure}

New observational programs on the GBT and WSRT of the well-studied Coma cluster have uncovered some exciting new phenomena.  The first is the discovery that the western edge of the halo is formed by a shock structure as shown in Figure \ref{comaedge}.  The shock nature of this feature is confirmed by the coincident sharp jump in X-rays \citep{markevitch} and the temperature jump by a factor of $\sim$3 in a reanalysis of the XMM data of \cite{comax}, as cited in \cite{browncoma}. 

This is a surprise. It could, of course, simply be a coincidence that an ordinary outgoing merger shock is found at the outer edge of a halo.   However, the halo at 330~MHz blends smoothly into the shock structure, which is why it has been hard to detect.  More likely, and more interesting, is that the halo and shock are physically connected,  \cite{markevitch} suggests that there may be a number of other examples of shocks at halo edges.  At present, halos are not thought to be caused by a single cluster-wide shock structure, and so the denouement of this issue will undoubtedly give us insights into halo formation.
We note that a detached, filamentary feature is also seen at the Coma halo shock location by \cite{pizzocoma} in WSRT observations at 150~MHz. 

On the GBT, we found that the relic to the southwest of the Coma cluster (first noticed by \cite{jaffe} but identified by \cite{comarelic}), was approximately twice as long as previously seen.   The new extent of the relic $\sim$2~Mpc, is not outside the norm.  However, we found that the relic is adjacent to a ``wall" of galaxies, the end of a 500 km/s infalling column previously seen as individual clumps and groups of infalling galaxies  \citep{infall}. Amazingly, this structure can be seen in the distribution of nebulae  by  \cite{wolf}. The interesting thing about this juxtaposition of the optical and radio structures is the likely dynamical state.   Since the column of galaxies has not yet merged with Coma, the relic cannot be a ``classical" outgoing shock.  However, in this early stage of the encounter,  the column can generate shocks at its leading and trailing edges; we see the latter.  We have suggested designating such features as ``infall shocks", rather than accretion shocks (as already used in the literature) or outgoing merger shocks, because their energetics and dynamical evolution will be quite distinct.
\begin{figure}[]
\resizebox{\hsize}{!}{\includegraphics[clip=true]{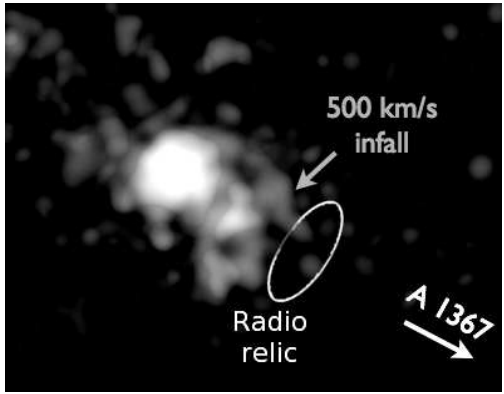}}
\caption{
\footnotesize
Optical galaxies and radio relic around the Coma cluster \citep{browncoma}.
}
\label{infall}
\end{figure}
\begin{figure}[]
\resizebox{\hsize}{!}{\includegraphics[clip=true]{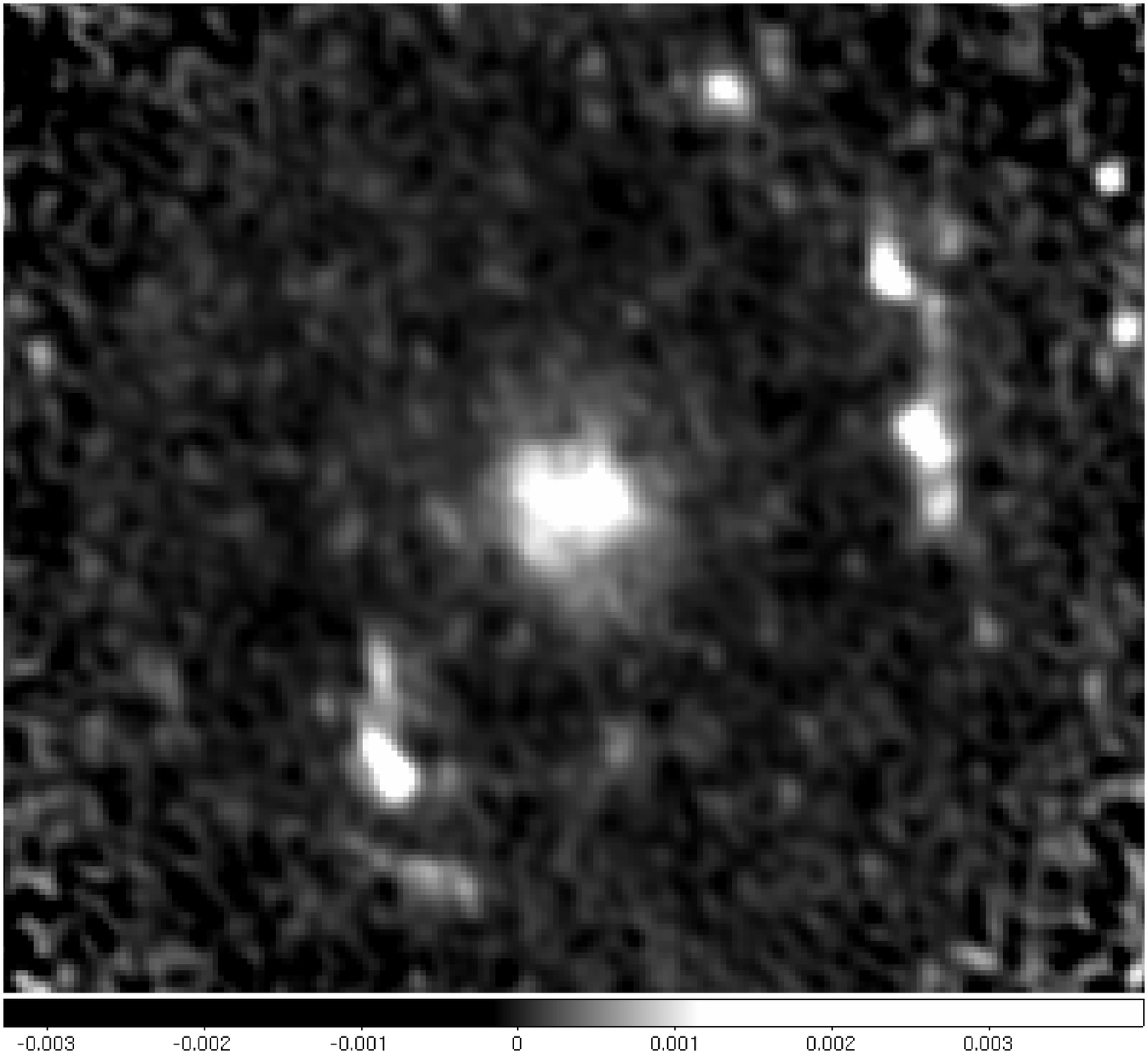}}
\caption{
\footnotesize
VLA 1.4 GHz image of cluster of galaxies at 0217+70 \citep{brown0217}  }
\label{0217}
\end{figure}
\subsection{A radio-identified cluster of galaxies}
One of the diffuse features found through our filtering work turned out to be a previously unidentified cluster of galaxies \citep{brown0217} at the low galactic latitude of 9$^o$, as seen in Figure \ref{0217}.  It has both a halo and double peripheral relics, seen as filamentary structures on a circle of  $\sim$1~Mpc radius around the cluster core.  Of special interest are the other filamentary structures seen with various orientations and at various positions around the halo and relic.   One possible explanation for these is that they are projections of the three-dimensional structures associated with merger shocks \citep{relicgrid}. However, the brightest examples are likely to be those seen projected along the line of sight, biasing our samples.   In the near future,  we shoulbe be able to get a much better picture of their 3D structure.

\begin{figure*}[t!]
\resizebox{\hsize}{!}{\includegraphics[clip=true]{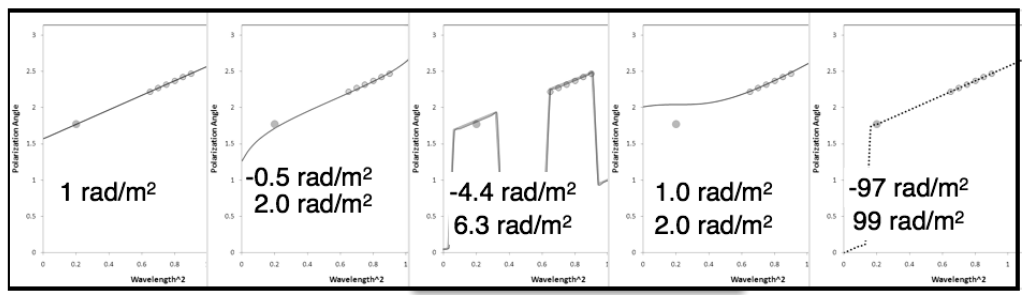}}
\caption{\footnotesize
Alternative models to describe the same (model) rotation measure observations.
}
\label{rm}
\end{figure*}

\section{Some shocking results on rotation measures}

We close with a brief note on some recent work on ambiguities in the determination of rotation measures \citep{farnsworth}.   The problem is illustrated in Figure \ref{rm}. The left-most image shows a model of observations with quite good coverage in $\lambda^2$ space, of a source with a rotation measure of 1~$\frac{rad}{m^2}$.  The other panels show a variety of two component models, all of them doing an excellent job of matching the observations at long wavelengths, and  two of the four matching the observations at both short and long wavelengths.   To date, most rotation measure data are much sparser than shown in these models;  there is a corresponding enormous growth in the number of alternative fits to the data.   Rotation measure synthesis \citep{rmsynth} can sometimes show the existence of two components, but fails in the very important case where two components are within the broad FWHM of the rotation measure transfer function.  There is {\em not} a simple prescription, such as the ratio $\Delta$RM/FWHM to tell whether an ambiguity will be present, since it also depends on the relative amplitudes and even intrinsic position angles of the two components.

The effects of these ambiguities on the physical properties derived from rotation measures are still unclear.   As a rough rule of thumb, RM studies of foreground screens will have an additional source of scatter, but no bias, because of these ambiguities.  Higher order RM studies, such as structure function analyses, may be seriously compromised at the smallest sampled scales.

Fortunately, there is a reasonable way to avoid these ambiguities.   The various models in Figure \ref{rm}, and multiple component models more generally, can be identified by examining the behavior of fractional polarization as a function of $\lambda^2$, or, even better, by examining fits to Q($\lambda^2$) and U($\lambda^2$).  Rotation Measures should {\em never} be determined using polarization angles alone, nor from RM synthesis alone.

\section{Conclusions}
There are three main take-home messages:  {\em 1)} Halos may be intimately connected with large scale shocks; {\em 2)} We are starting to see, and need to carefully distinguish, between a wide variety of cluster-related shocks; and {\em 3) }Verify Q \& U behavior when determining RMs.

\begin{acknowledgements}
 Former Minnesota undergraduates Jon Duesterhoeft and Jeff Lemmerman made important contributions to these efforts  This work is partially supported at the University of Minnesota by U.S. NSF grant AST 0908688.  The cited papers detail support for the VLA and WSRT observatories, on which much of this work is based.  \end{acknowledgements}

\bibliographystyle{aa}

\end{document}